\documentstyle{mn}
\input epsf
\epsfverbosetrue
\begin{document}
\title[Nonlinear spherical Alfv\'en waves]{Nonlinear spherical Alfv\'en waves}

\author[U. Torkelsson \& G. C Boynton]{Ulf Torkelsson,$^1$ 
G. Christopher Boynton,$^2$\\
$^1$Institute of Astronomy, Madingley Road, Cambridge CB3 0HA,
United Kingdom\\
$^2$Physics Department, Univ. of Miami, P. O. Box 248046, Coral Gables,
FL 33124, USA}

\maketitle

\begin{abstract}
We present an one-dimensional numerical study of Alfv\'en waves propagating
along a radial magnetic field.
Neglecting losses,
any spherical Alfv\'en wave, no matter how small its initial
amplitude is, becomes nonlinear at sufficiently large radii.  From
previous simulations of Alfv\'en waves in plane parallel atmospheres we did
expect the waves to steepen and produce current sheets in the nonlinear regime,
which was confirmed by our new calculations.  On the other hand we did find
that even the least nonlinear waves were damped out almost completely before 10 
$R_{\sun}$.
A damping of that kind
is required by models of Alfv\'en wave-driven winds from old low-mass stars
as these winds are mainly accelerated within a few stellar radii.
\end{abstract}

\begin{keywords}
MHD -- waves -- solar wind -- stars: mass-loss
\end{keywords}

\section{Introduction}

Spherically symmetric models are often the simplest realistic models conceivable
for astrophysical systems, and have been applied successfully to describe not 
only hydrostatic objects such as stars, but also hydrodynamic processes, for
instance stellar winds (Parker 1958).
In many cases it is
desirable to extend a spherically symmetric hydrodynamic problem to the
analogous magnetohydrodynamic problem.  In a global sense such a 
magnetohydrodynamic analogue cannot exist, as that would require the 
existence of 
magnetic monopoles.  Locally, however, the magnetic field is divergence-free,
and presents a natural approximation of a region with a diverging magnetic
field.  In hydrodynamics there are only acoustic waves (for spherical sound
waves see e.g. Landau \& Lifshitz 1987), but the magnetic field
introduces Alfv\'en waves.  These transverse magnetic field oscillations 
cannot be
extended to cover an entire spherical surface without introducing 
discontinuities, but once again it will not present any difficulties locally.

In a previous paper (Boynton \& Torkelsson 1996, hereafter Paper 1) we have 
shown that nonlinear Alfv\'en waves in a planar geometry can steepen and form
current sheets, and thereby be damped by Joule dissipation or by doing
mechanical work on the background medium.
The efficiency of this mechanism is limited by the Alfv\'en wave becoming
less nonlinear as it propagates upwards through a stratified medium.  A 
spherical 
Alfv\'en wave on the other hand may become less nonlinear for some time, but
eventually has to grow nonlinear again because of the divergence of
the background magnetic field.
Furthermore the gas pressure decreases faster than the magnetic pressure with
height independently of whether the symmetry is plane-parallel or spherical,
so that the Alfv\'en wave becomes more dynamically important.
Models of Alfv\'en wave-driven outflows have been constructed by Hartmann \& 
MacGregor (1980, 1982) to explain the winds from late-type giants, although 
a more popular model is that the outflows are driven by radiation pressure 
working on dust in the stellar atmosphere.  It has been pointed out that the
Alfv\'en waves must damp within a few stellar radii to avoid accelerating the
wind to too high velocities.  There is no generally accepted model for this
damping and Holzer, Fl\aa ~\& Leer  (1983) have pointed out that the stellar 
wind
depends sensitively on the damping mechanism.

A spherically symmetric model may also apply to the coronal holes observed
on the Sun (Bohlin 1976, Zirker 1977).  The magnetic field in a coronal
hole is open, which allows the plasma to expand outwards and form
the so-called fast solar wind.  This component of the solar wind is too
fast to be described by the classical Parker (1958) model, and it is possible
that the extra acceleration is provided by Alfv\'en waves.  Alfv\'en waves 
have been observed further out in the solar wind (Belcher \& Davis 1971, Balogh
et al. 1995), but
it is not known how these waves relate to the ones that are supposedly present
in the corona.

Section 2 describes the initial hydrostatic model and summarizes the properties
of linear waves propagating through it.  
The results of our numerical simulations
are presented in Sect. 3.
Section 4 discusses the physical interpretation and the astrophysical 
consequences of our 
results, in particular, with respect to solar and stellar winds, and our 
conclusions are summarized in Sect. 5.

\section{The static model and its wave modes}
\subsection{The static model}

\begin{figure}
\epsfxsize=8.8cm
\epsfbox{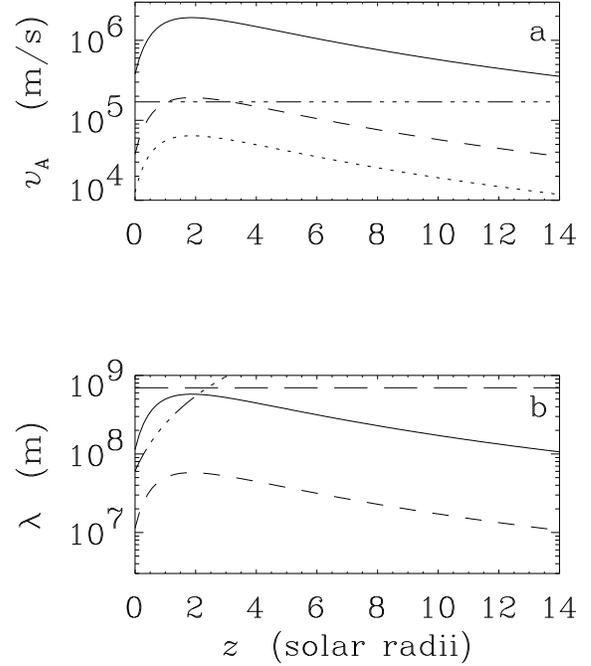}
\caption{({\bf a})  Alfv\'en velocity as a function of height, $z$, 
for magnetic field 
strengths of $3\,10^{-4}$ T (solid line), $3\,10^{-5}$ T (dashed line) and
$1\,10^{-5}$ T (dotted line) as measured at $z = 0$.  The sound speed is 
plotted as a dash-dotted
line.
({\bf b})  The wavelength of an Alfv\'en wave with a period of 300 s as a 
function of $z$ for the same magnetic field strengths.  We do not
plot the wavelength for $1\,10^{-5}$ T, as its period of 900 s
gives it the same wavelength as $3\,10^{-5}$
T with a period of 300 s.
The dash-dotted line denotes the local pressure scale height
and the long-dashed line the solar radius}
\label{wave_length}
\end{figure}

We assume a spherically symmetric Sun-like star.
On top of the
stellar surface there is an isothermal corona at a temperature of $10^6$ K.
For a corona in hydrostatic equilibrium the density can be written
as
\begin{equation}
  \rho_0\left(z\right) = \rho_0\left(0\right) 
\exp\left(- \frac{R}{H}\frac{z}{R+z}\right),
\label{stratification}
\end{equation}
where $\rho_0\left(0\right)$ is the coronal density at the stellar surface, 
$R$ is the
stellar radius, $z$ the height above the stellar surface and $H =
2 k_{\rm B} TR^2/(GMm_{\rm H}) = 6.1\,10^7$\,m, a scale height, with 
$k_{\rm B}$ Boltzmann's constant,
$T$ the temperature, $G$ the gravitational constant, $M$ the mass of 
the star, and $m_{\rm H}$ the mass of a hydrogen atom.  
Throughout this paper subscript 0s, with a few obvious exceptions,
refer to the undisturbed background and 0s as
arguments to functions mean the functional values at $z = 0$, that is the
stellar surface.
The 
isothermal sound speed, $c_{\rm s}$, which is what matters as we are 
assuming
isothermality in the dynamical model, is $1.3\,10^5$ m\,s$^{-1}$ and 
$\rho_0(0)$ is put to
$5\,10^{-13}$ kg\,m$^{-3}$.  
A vertical magnetic field, $B_z(z) \propto (R+z)^{-2}$,
gives an Alfv\'en speed, $v_{\rm A} = B_z/\sqrt{\mu_0 \rho}$, with the
$z$-dependence of Fig. \ref{wave_length}a.
For $B_z(0) = 3\,10^{-4}$ T, the
Alfv\'en wave is supersonic even at $z = 14\,R_{\sun}$, 
whereas for 
$B_z(0) = 10^{-5}$ T the Alfv\'en wave is subsonic everywhere.
Assuming a period of 300 s
the wavelength is always small compared to the scale height 
(Fig. \ref{wave_length}b), except for
$B_z(0) = 3\,10^{-4}$\,T when it exceeds the 
scale height for $z < 2 R_{\sun}$ and is comparable 
to the solar radius at $z \approx 1.5 R_{\sun}$.  
Note that for $B_z(0) = 10^{-5}$ T we choose a period of 900 s, to get the
same wavelength as for $B_z(0) = 3\,10^{-5}$ T.

\subsection{Linear theory of spherical waves}

The magnetohydrodynamic equations for a spherically symmetric system can be 
written as
\begin{equation}
  \frac{\partial \rho}{\partial t} + \frac{1}{\left(R+z\right)^2}
\frac{\partial}{\partial z}\left[\left(R+z\right)^2 \rho v_z\right] = 0,
\label{mass}
\end{equation}
\begin{eqnarray}
  \frac{\partial}{\partial t}\left(\rho v_x\right) + 
\frac{1}{\left(R+z\right)^2} \frac{\partial}{\partial z} \left[\left(R+z
\right)^2 \rho v_x v_z\right] = \nonumber \\
B_z J_y - \rho \frac{v_x v_z}{R+z},
\label{vx}
\end{eqnarray}
\begin{eqnarray}
  \frac{\partial}{\partial t} \left(\rho v_y \right) +
\frac{1}{\left(R+z\right)^2} \frac{\partial}{\partial z} \left[\left(R+z
\right)^2 \rho v_y v_z\right] = \nonumber \\
- B_z J_x - \rho \frac{v_y v_z}{R+z},
\label{vy}
\end{eqnarray}
\begin{eqnarray}
  \frac{\partial}{\partial t} \left(\rho v_z\right) + 
\frac{1}{\left(R+z\right)^2} 
\frac{\partial}{\partial z} \left[\left(R+z\right)^2 \rho v_z^2\right] =
\nonumber \\
- \frac{\partial p}{\partial z} - B_x J_y + B_y J_x - \rho g + \rho 
\frac{v_x^2 + v_y^2}{R+z},
\label{vz}
\end{eqnarray}
\begin{eqnarray}
  \frac{\partial B_x}{\partial t} + \frac{1}{R+z} \frac{\partial}{\partial z}
\left[\left(R+z\right) B_x v_z\right] = \nonumber \\
\frac{1}{R+z} \frac{\partial}
{\partial z} \left[\left(R+z\right) B_z v_x\right],
\label{bx}
\end{eqnarray}
\begin{eqnarray}
  \frac{\partial B_y}{\partial t} + \frac{1}{R+z} \frac{\partial}{\partial z}
\left[\left(R+z\right)B_y v_z\right] = \nonumber \\
\frac{1}{R+z}
\frac{\partial}{\partial z} \left[\left(R+z\right) B_z v_y\right],
\label{by}
\end{eqnarray}
where $\rho$, $v_x$, $v_y$, $B_x$ and $B_y$ stand for the density and the 
transverse components of
the velocity and the magnetic field.  $v_z$ is the velocity along the
background magnetic field, $g$ the gravitational field strength,
and $p = c_{\rm s}^2 \rho$ is the pressure.
The need for an energy 
equation is eliminated by assuming all processes to be isothermal, and $B_z$
must be independent of time to keep the magnetic field divergence-free.
The electric current density is given by
\begin{equation}
J_x = - \frac{1}{\mu_0} \frac{1}{R+z} \frac{\partial}{\partial z} 
\left[\left(R+z\right)B_y\right],
\end{equation}
and
\begin{equation}
J_y = \frac{1}{\mu_0} \frac{1}{R+z}
\frac{\partial}{\partial z}\left[\left(R+z\right) B_x\right],
\end{equation}
where $\mu_0$ is the permeability of free space.  

We linearize the equations around a, possibly, stratified static
background medium
and a radial magnetic field $B_z = B_z(0) \left(\frac{R}{R+z}\right)^2$. 
For brevity we leave out the $y$-components, and use $r = R+z$.
\begin{equation}
  \frac{\partial \tilde \rho}{\partial t} + \frac{1}{r^2} \frac{\partial}
{\partial r} \left(r^2 \rho_0 v_z\right) = 0,
\label{lin_mass}
\end{equation}
\begin{equation}
\frac{\partial}{\partial t} \left(\rho_0 v_x\right) = B_z(0) 
\left(\frac{R}{r}\right)^2 \frac{1}{\mu_0}\frac{1}{r} 
\frac{\partial}{\partial r}\left(rB_x\right),
\label{lin_vx}
\end{equation}
\begin{equation}
  \frac{\partial}{\partial t} \left(\rho_0 v_z\right) = - c_{\rm s}^2
\frac{\partial \tilde \rho}{\partial r} - \tilde \rho g,
\label{lin_vz}
\end{equation}
\begin{equation}
  \frac{\partial B_x}{\partial t} = \frac{1}{r} \frac{\partial}{\partial r}
\left(r B_z(0) \left(\frac{R}{r}\right)^2 v_x\right),
\label{lin_bx}
\end{equation}
where we have written the density as $\rho = \rho_0 + \tilde \rho$.
These equations separate into two groups, Eqs. (\ref{lin_mass}) and
(\ref{lin_vz}) describing acoustic waves, and Eqs. (\ref{lin_vx}) and 
(\ref{lin_bx}) describing Alfv\'en waves.  

As a first example  we assume that $\rho_0$ is constant and $g = 0$.  
The acoustic waves are then described by two wave equations
\begin{equation}
  \frac{\partial^2 \tilde \rho}{\partial t^2} = c_{\rm s}^2 \frac{1}{r^2}
\frac{\partial}{\partial r} \left(r^2 \frac{\partial \tilde \rho}{\partial r}
\right),
\end{equation}
and
\begin{equation}
  \frac{\partial^2 v_z}{\partial t^2} = c_{\rm s}^2 \frac{\partial}{\partial r}
\left[ \frac{1}{r^2} \frac{\partial}{\partial r} \left(r^2 v_z\right)\right],
\end{equation}
which have the solutions 
\begin{equation}
  \tilde \rho = \tilde \rho\left(0\right) \frac{R}{r} \mbox{e}^{i
\left(kr -\omega t\right)},
\end{equation}
where $\tilde \rho(0)$ is the amplitude of the density fluctuations at $r = R$
and
\begin{equation}
  v_z = \omega \frac{\tilde \rho\left(0\right)}
{\rho_0} R \frac{\partial}{\partial r} 
\left(\frac{\mbox{e}^{i\left(kr-\omega t\right)}}{ik^2r}\right)
\end{equation}
with the dispersion relation
\begin{equation}
  \omega^2 = c_{\rm s}^2 k^2
\end{equation}
(cf. Landau \& Lifshitz 1987, Ch. 70).

\begin{figure}
\epsfxsize=8.8cm
\epsfbox{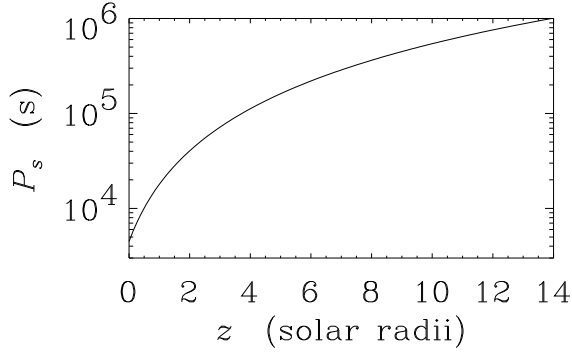}
\caption{The maximal period for acoustic waves in a stratified
atmosphere, $P_{\rm s}$, as a function of $z$}
\label{cutoff}
\end{figure}

  In the case of a stratified medium the wave equation for the sound waves
can be written as
\begin{equation}
  \frac{\partial^2 \tilde \rho}{\partial t^2} = \frac{c_{\rm s}^2}{r^2}
\frac{\partial}{\partial r} \left(r^2 \frac{\partial \tilde \rho}{\partial r}
\right) + g \frac{\partial \tilde \rho}{\partial r}.
\end{equation}
For $g = GM/r^2$ the background density is $\propto \exp\left(- \frac{R}{H}
\frac{z}{R+z}\right)$, where $H$ is the scale height at $z = 0$.  We write the
density fluctuations as 
\begin{equation}
  \tilde \rho = \tilde \rho\left(0\right) \frac{R}{r} \exp\left(- \frac{R}{2H} 
\frac{z}{R+z}\right) \mbox{e}^{i\left(kz - \omega t\right)}
\end{equation}
(cf. Paper 1).
The resulting dispersion relation is
\begin{equation}
  \omega^2 = c_{\rm s}^2 k^2 + \frac{c_{\rm s}^2 R^4}{4 H^2 r^4},
\end{equation}
implying that acoustic waves with frequencies $\omega < N_{\rm s} =
c_{\rm s}R^2/(2 H r^2)$ are evanescent (cf. Lamb 1908, 1932).
We plot the corresponding period, $P_{\rm s} = 2\pi/N_{\rm s}$ in Fig. 
\ref{cutoff}.

\begin{figure}
\epsfxsize=8.8cm
\epsfbox{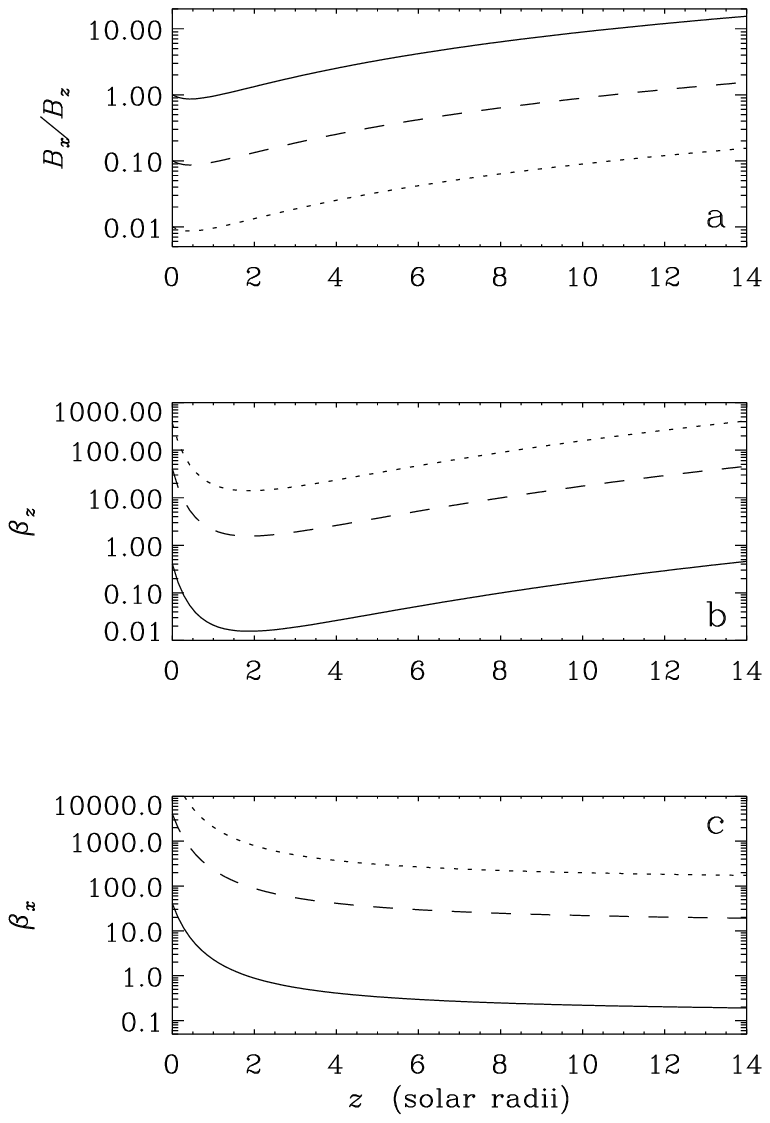}
\caption{({\bf a}) The relative amplitude of an Alfv\'en wave propagating 
through a stratified atmosphere.  The solid, dashed and dotted lines represent
waves of initial relative amplitudes of 1, 0.1 and 0.01, respectively.  
({\bf b}, 
{\bf c}) The ratio of gas to magnetic pressure for the background field, $B_z$
({\bf b}) and the oscillating field $B_x$ ({\bf c}).  
$B_z$ is $3\,10^{-4}$ T (solid line), 
$3\,10^{-5}$ T (dashed line), and $10^{-5}$ T dotted line.  The relative 
amplitude of the oscillations is $B_x(0)/B_z(0) = 0.1$}
\label{beta}
\end{figure}

\begin{table*}
\caption{Simulations of linearly polarized Alfv\'en waves with period, $P$,
travelling through
an isothermal medium stratified according to Eq. (\ref{stratification}) with a
base density $5\,10^{-13}$ kg\,m$^{-3}$ and 
a background magnetic field, $B_z$, decreasing as 
$1/(R+z)^2$.  The temperature of the medium is $10^6$ K, which corresponds
to a sound speed of $1.29\,10^5$ m\,s$^{-1}$.  
The computational domain covers the
the corona between 1 and 15 $R_{\sun}$ outside of a solarlike star for Models
1a - 1c, and between 1 and 16 $R_{\sun}$ for Models 2a - 3c.
The homogeneous (H) models are similar but has a constant density
of $10^{-14}$ kg\,m$^{-3}$, their radial extents are 0.15 $R_{\sun}$ (H1a-H1c),
0.3 $R_{\sun}$ (H1d)
and 0.6 $R_{\sun}$ (H2a-H2c) with the lower boundary at $1 R_{\sun}$.
The driven Alfv\'en wave is polarized at $45\degr$ 
to the $x$-axes for all of the simulations.
$\Delta t$ denotes the time step and $N$ the number
of spatial grid points used in the respective models.  $B_z(0)$ is the strength
of the background magnetic field, $v_{\rm A}(0)$ the Alfv\'en velocity,
$(B_{\rm osc}/B_z)(0)$ the relative amplitude of
the Alfv\'en wave
and $\beta_{\rm mag}(0)$ the plasma beta all calculated at $z = 0$} 
\label{lin_pol}
\begin{tabular}{llrlrrrr}
\hline \\
Model & $\Delta t$ (s) & $N$ & $B_z(0)$ (T) & $P$ (s) & 
$v_{\rm A}(0)$ (m\,s$^{-1}$) & 
$(B_{\rm osc}/B_z)(0)$ & $\beta_{\rm mag}(0)$ \\
\hline \\
H1a & 0.25 & 720 & $5\,10^{-6}$ & 300 & $4.4\,10^4$ & 0.01 & 17 \\
H1b & 0.25 & 720 & $5\,10^{-6}$ & 300 & $4.4\,10^4$ & 0.1 & 17 \\
H1c & 0.25 & 720 & $5\,10^{-6}$ & 300 & $4.4\,10^4$ & 1.0 & 17 \\
H1d & 0.25 & 1440 & $5\,10^{-6}$ & 300 & $4.4\,10^4$ & 1.0 & 17 \\
H2a & 1 & 720 & $2\,10^{-5}$ & 300 & $1.8\,10^5$ & 0.01 & 1.0 \\
H2b & 1 & 720 & $2\,10^{-5}$ & 300 & $1.8\,10^5$ & 0.1 & 1.0 \\
H2c & 1 & 720 & $2\,10^{-5}$ & 300 & $1.8\,10^5$ & 1.0 & 1.0 \\
1a & 1 & 15120 & $1\,10^{-5}$ & 900 & $1.3\,10^4$ & 0.01 & 210 \\
1b & 1 & 15120 & $1\,10^{-5}$ & 900 & $1.3\,10^4$ & 0.1 & 210 \\
1c & 1 & 15120 & $1\,10^{-5}$ & 900 & $1.3\,10^4$ & 1.0 & 210 \\
2a & 1 & 15120 & $3\,10^{-5}$ & 300 & $3.8\,10^4$ & 0.01 & 23 \\
2b & 1 & 15120 & $3\,10^{-5}$ & 300 & $3.8\,10^4$ &0.1 & 23 \\
2c & 1 & 15120 & $3\,10^{-5}$ & 300 & $3.8\,10^4$ &1.0 & 23 \\
3a & 0.25 & 7560 & $3\,10^{-4}$ & 300 & $3.8\,10^5$ &0.01 & 0.23 \\
3b & 0.25 & 7560 & $3\,10^{-4}$ & 300 & $3.8\,10^5$ & 0.1 & 0.23 \\
3c & 0.25 & 7560 & $3\,10^{-4}$ & 300 & $3.8\,10^5$ &1.0 & 0.23 \\
\hline \\
\end{tabular}
\end{table*}

Equations (\ref{lin_vx}) and (\ref{lin_bx}) can be rewritten as
(cf. Heinemann \& Olbert 1980, Leer, Holzer \& Fl\aa ~1982, MacGregor \&
Charbonneau 1994)
\begin{equation}
  \frac{\partial f}{\partial t} = v_{\rm A}\left(0\right) 
\left(\frac{R}{r}\right)^2 
\left(\frac{g}{r} - \frac{\partial f}{\partial r}\right),
\end{equation}
and
\begin{equation}
  \frac{\partial g}{\partial t} = v_{\rm A}\left(0\right) 
\left(\frac{R}{r}\right)^2
\left(-\frac{f}{r} + \frac{\partial g}{\partial r}\right),
\end{equation}
where
\begin{equation}
  f = v_x - \frac{B_x}{\sqrt{\mu_0 \rho_0}},\,\,\, g = v_x + 
\frac{B_x}{\sqrt{\mu_0 \rho_0}},
\end{equation} 
and $v_{\rm A}(0)
= B_z(0)/\sqrt{\mu_0 \rho_0(0)}$ is the Alfv\'en velocity at $r = R$, so that
\begin{equation}
  v_x = \frac{1}{2}\left(f + g\right)
\end{equation}
and
\begin{equation}
  B_x = \frac{1}{2} \sqrt{\mu_0\rho_0} \left(- f + g\right).
\end{equation}
Note that $ f = 0$ and $g = 0$
yield waves propagating downwards and upwards, respectively. 
Substituting $f(r,t) = F(r)\mbox{e}^{-i\omega t}$ and $g(r,t) = G(r) 
\mbox{e}^{-i\omega t}$ we get
\begin{equation}
\frac{\mbox{d}}{\mbox{d}r}
\left(\begin{array}{c}
F \\
\hfill \\
G
\end{array}\right)
= 
\left(\begin{array}{cc}
\frac{i\omega}{v_{\rm A}}\left(\frac{r}{R}\right)^2 & \frac{1}{r} \\
&\hfill \\
\frac{1}{r} & - \frac{i\omega}{v_{\rm A}} \left(\frac{r}{R}\right)^2 
\end{array}\right)
\left(\begin{array}{c}
F \\
\hfill \\
G
\end{array}\right)
\end{equation}
The general solution to these equations is rather complicated, and we will
restrict ourselves to cite some general and useful results, that can be 
derived from simple physical arguments.

First we derive the scaling properties of the amplitudes of $v_x$ and $B_x$ in
the WKB sense.
Consider the Poynting flux
\begin{equation}
  {\bmath S} = \frac{{\bmath E \times \bmath B}}{\mu_0},
\end{equation}
with the electric field, ${\bmath E} = - {\bmath v \times \bmath B} + 
{\bmath J}/\sigma$, 
where
we assume that the conductivity, $\sigma$, is infinite.
If there are no losses, the amplitude of the radial component of the 
Poynting vector $S_{z}$ 
will be proportional
to $1/r^2$.  $S_{z}$
is also proportional to $B_z B_x v_x$, the oscillating 
transverse magnetic 
and velocity fields, and as 
$B_z \propto 1/r^2$ in our
monopole geometry, $B_x v_x$ is not explicitly dependent upon 
$r$.
The oscillating magnetic and velocity fields of an outgoing Alfv\'en wave are 
related by 
\begin{equation}
  v_x = - \frac{B_x}{\sqrt{\mu_0\rho_0}},
\end{equation}
which gives $B_x \propto 
\rho_0^{1/4}$ and $v_x \propto \rho_0^{-1/4}$.
We plot the relative amplitude of the Alfv\'en wave, $B_x/B_z$, in Fig. 
\ref{beta}a.  In Figs. \ref{beta}b and c we compare $B_z$ and $B_x$, 
respectively, to the gas pressure by calculating the magnetic beta $\beta_{x,z}
= 2 \mu_0 p(z)/B_{x,z}^2$.

The main effect of the deviations from the WKB-approximation is that the waves
can be reflected against gradients in the Alfv\'en speed.  This problem has 
been discussed for linearized spherical Alfv\'en waves by An et al. (1990) and
Lou \& Rosner (1994).  The reflection is important as it increases the momentum
transferred from the Alfv\'en wave to the medium.

\section{Results of the numerical simulations}

We modified the code of Paper 1 to simulate Alfv\'en waves in a spherically
symmetric magnetic field.
In addition we now solve for
both transverse components of the velocity and magnetic fields, so that
also circularly polarized waves can be described.
In general our waves are linearly polarized at $45\degr$ to the $x$-axis.
Initial test runs showed that the boundary conditions from
Paper 1 were too reflective at the upper boundary.  After some experiments we
found that a satisfactory solution was to extrapolate the boundary values of
$\rho v_z$, $B_x$ and $B_y$ from the last three grid points, and 
calculate $\rho v_x$ and $\rho v_y$ from the requirement that they should
describe Alfv\'en waves propagating out through the boundary.
The boundary conditions at the lower boundary are the same as in Paper 1, that
is we drive an upwards propagating Alfv\'en wave, and keep all other variables
fixed at their initial values.

Table \ref{lin_pol} describes the models that we have calculated.  The H-models 
are homogeneous in the sense that they have constant density and temperature, 
and no gravity,
but the magnetic field still goes as $1/(R+z)^2$.

\begin{figure}
\epsfxsize=8.8cm
\epsfbox{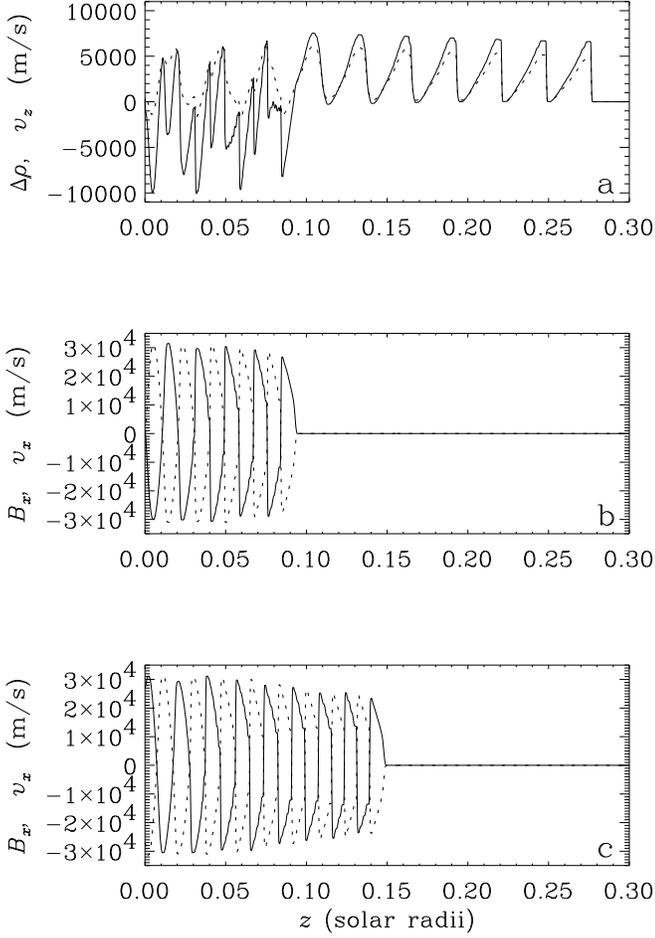}
\caption{Magnetohydrodynamic waves propagating in a spherically symmetric
magnetic field with a uniform background density (Model H1d).  
({\bf a})  $\Delta \rho$ (solid
line), and $v_z$ (dotted line)
oscillations at 6\,000 s.  
Both quantities 
are measured in velocity units.  The lower two panels show
$B_x$ (solid line)
and $v_x$ (dotted line) oscillations at ({\bf b}) 6\,000 s and
({\bf c}) 10\,000 s, respectively, both measured
in velocity units.
({\bf a}) shows the acoustic precursor ($z > 0.1 R_{\sun}$) to the Alfv\'en 
wave
in ({\bf b}), and the oscillations generated by the magnetic pressure 
($z < 0.1 R_{\sun}$).  
The Alfv\'en wave is gradually damped as it steepens 
into current sheets ({\bf c})}
\label{precursor}
\end{figure}

\begin{figure}
\epsfxsize=8.cm
\epsfbox{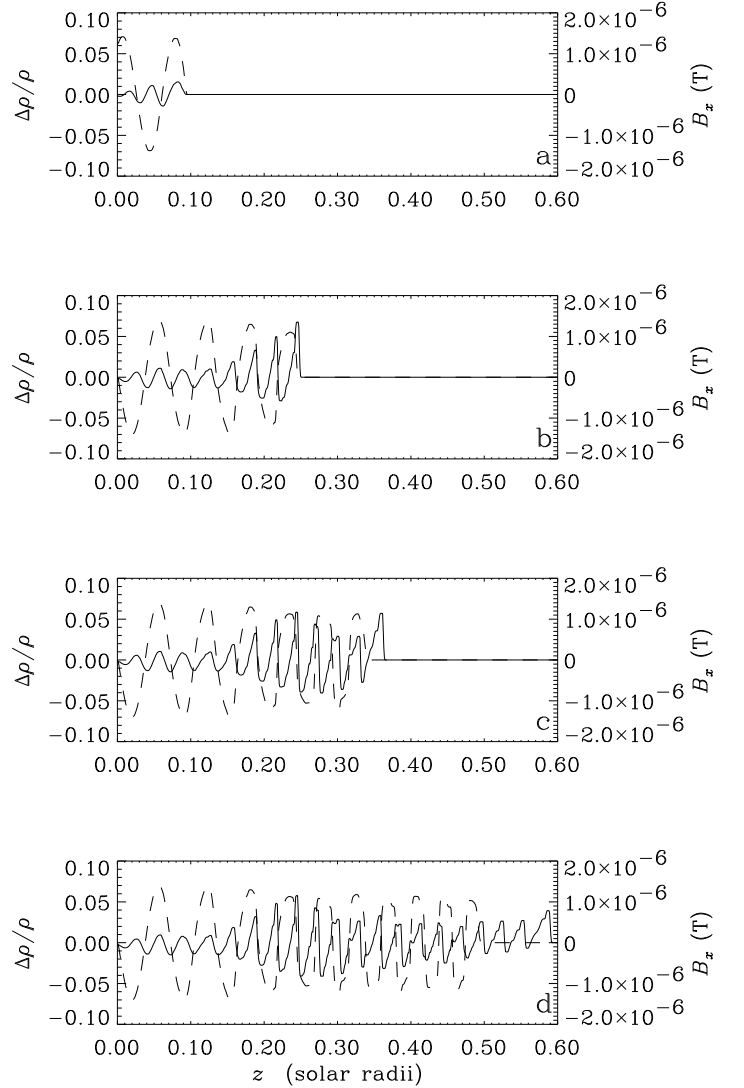}
\caption{Magnetohydrodynamic waves propagating in a spherically symmetric
magnetic field with a uniform background density (Model H2b).  The left
axis and the solid line show $\Delta \rho/\rho_0$
and the right axis and the dotted line $B_x$.
The times are ({\bf a}) 400 s, ({\bf b}) 1\,200 s, ({\bf c}) 1\,800 s, and
({\bf d}) 3\,000 s, respectively.  At the start the Alfv\'en wave is propagating
faster than a sound wave so that in ({\bf a}) we see density oscillations
carried by the magnetic field, but as the Alfv\'en velocity decreases there is
a sound 
wave that overtakes the Alfv\'en wave ({\bf b}-{\bf d})}
\label{precursor_1}
\end{figure}

\begin{table}
\caption{The maximum of the precursory density fluctuations 
$\left(\Delta \rho/\rho_0 \right)_{1{\rm max}}$ and of the second-order 
fluctuations
$\left(\Delta \rho/\rho_0 \right)_{2{\rm min}}$ in the interacting Alfv\'en and
acoustic waves for Models H1a-H1c}
\label{amp_scale}
\begin{tabular}{lrr}
\hline \\
$(B_{\rm osc}/B_z)(0)$ & 
$\left(\Delta \rho/\rho_0 \right)_{1{\rm max}}$ &
$\left(\Delta \rho/\rho_0 \right)_{2{\rm min}}$ \\
\hline \\
0.01 & $4.8\,10^{-6}$ & $-6\,10^{-6}$ \\
0.1 & $4.8\,10^{-4}$ & $-6\,10^{-4}$ \\
1.0 & $4.7\,10^{-2}$ & $-6\,10^{-2}$ \\
\hline \\
\end{tabular}
\end{table}

\begin{figure}
\epsfxsize=8.8cm
\epsfbox{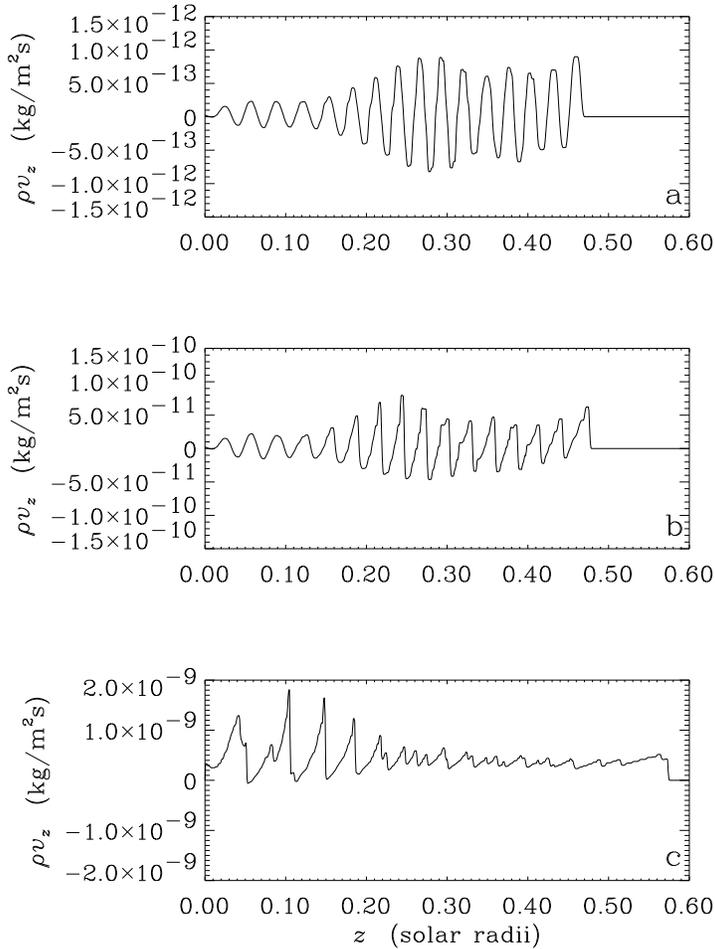}
\caption{$\rho v_z$ as a function of $z$ for Models H2a - c ({\bf a} - {\bf c})
at 2\,400 s.  Note how $\rho v_z$ changes character from being a propagating
wave
({\bf a},{\bf b}) to an outflow ({\bf c}) when the amplitude of the Alfv\'en 
wave increases}
\label{outflow}
\end{figure}

\subsection{Wave propagation}

\subsubsection{Unstratified models}

The amplitudes of the transverse velocities and magnetic fields are not 
affected by the divergence of the radial magnetic field, but they are affected
by the stratification in the same way as in the plane-parallel case considered
in Paper 1.
The most straightforward cases are obviously the ones without stratification,
where the Alfv\'en velocity decreases as $1/(R+z)^2$.  Models H1a-d are 
particularly simple as $v_{\rm A} < c_{\rm s}$ everywhere in these models,
so that the Alfv\'en wave has an acoustic precursor.
For high amplitude waves, such as H1c and H1d, both the 
acoustic
precursor and the Alfv\'en wave steepen to form shocks and current sheets, 
respectively (Fig. \ref{precursor}). 
Note that we have transformed the density, $\Delta \rho =
\rho - \rho_0$, and magnetic field, $B_x$, 
oscillations to velocities by multiplying with $c_s/\rho_0$ and $1/\sqrt{\mu_0
\rho_0}$, respectively.  We quantify the amplitude of the precursor by 
defining $\left(\frac{\Delta \rho}{\rho_0}\right)_{1{\rm max}}$ as the
maximum
of the density fluctuations in the acoustic precursor (the density does not
drop below its background value).  Likewise we call the
minimum of the density fluctuations inside the Alfv\'en wave $\left(
\frac{\Delta \rho}{\rho_0} \right)_{2{\rm min}}$.  
The numerical values for these
quantities are given in Tab. \ref{amp_scale} for Models H1a - H1c.
Both the precursors and the density fluctuations inside the
Alfv\'en wave are proportional to the square of the amplitude of the Alfv\'en
wave.  This is natural as they are produced by the magnetic pressure of the
Alfv\'en wave.

The situation is more complicated in Models H2a-c, where $v_{\rm A} > 
c_{\rm s}$ at $z = 0$, but $c_{\rm s} > v_{\rm A}$ for $z > 0.04 R_{\sun}$.  
The acoustic
wave lags behind the Alfv\'en wave at first.  The density oscillations
propagating with the Alfv\'en wave in Fig \ref{precursor_1}a are not
acoustic waves, but rather
$\rho$ and $v_z$ oscillations caused by the magnetic pressure oscillations
(cf. Hollweg 1971).  Density maxima occur initially where there is a maximum 
in the
magnetic pressure, but drift out of phase when the acoustic
wave overtakes the Alfv\'en wave (Fig. \ref{precursor_1}b,c).  
The density fluctuations simultaneously increase in amplitude, and eventually
steepen into shocks and lose energy (Fig. \ref{precursor_1}d).  
It is instructive to study these oscillations as a function of the
amplitude of the Alfv\'en waves.  For a low amplitude Alfv\'en wave the
density fluctuations are essentially a propagating wave which does not
transport any significant amount of mass (Fig. 
\ref{outflow}a).  
This
wave is growing in amplitude up to $z = 0.3 R_{\sun}$ after
which it remains constant or even decreases in amplitude again (Fig. 
\ref{outflow}b).  The amplitude of the wave is varying roughly as the
square of the amplitude of the Alfv\'en wave as can be found by comparing 
Figs. \ref{outflow}a and b, and it is propagating at the sound speed,
$1.3\,10^5$ m\,s$^{-1}$.
$\rho v_z$ changes character 
completely for a nonlinear Alfv\'en wave (Fig. \ref{outflow}c).  There is
an outflow of mass, which
is concentrated to shocks that are separated in time by roughly
half the period of the Alfv\'en wave.  The front of the outflow is initially
propagating at $2\,10^5$ m\,s$^{-1}$, but slows down
to $1.5\,10^5$ m\,s$^{-1}$, at 3\,000 s, which is still supersonic.  However the
outflow velocity, $v_z$, is mainly subsonic, although it may reach peak values
as high as $1.4\,10^5$ m\,s$^{-1}$.

\begin{figure*}
\epsfxsize=18cm
\epsfbox{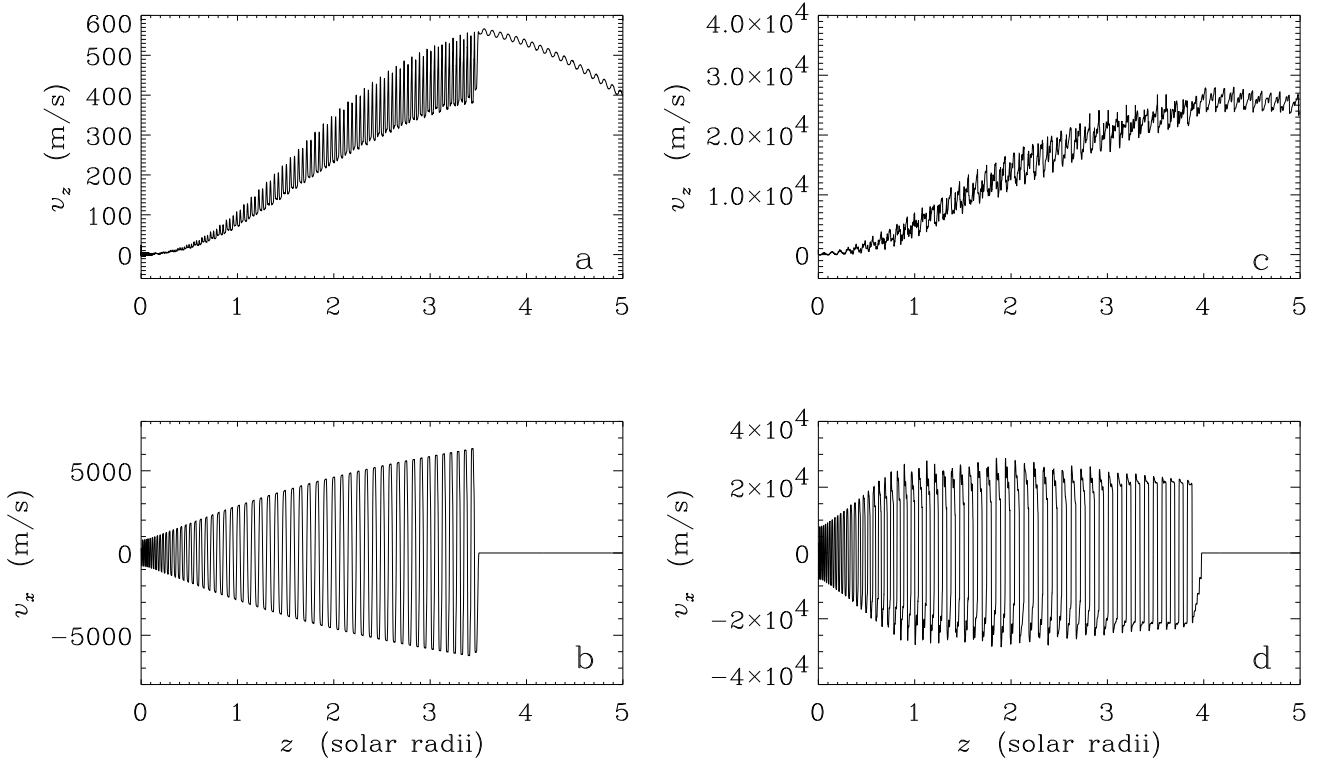}
\caption{$v_z$ and $v_x$ at 50\,000 s for Models 1b ({\bf a}) and ({\bf b}),
and 1c ({\bf c}) and ({\bf d}).  ({\bf a}, {\bf b}) show
an Alfv\'en wave of low amplitude with an acoustic precursor.  
At higher amplitude
({\bf c}, {\bf d}) the Alfv\'en wave is strongly damped by nonlinear 
steepening, and the Alfv\'en wave front is propagating faster}
\label{precursor_6a}
\end{figure*}

\begin{figure}
\epsfxsize=7.2cm
\epsfbox{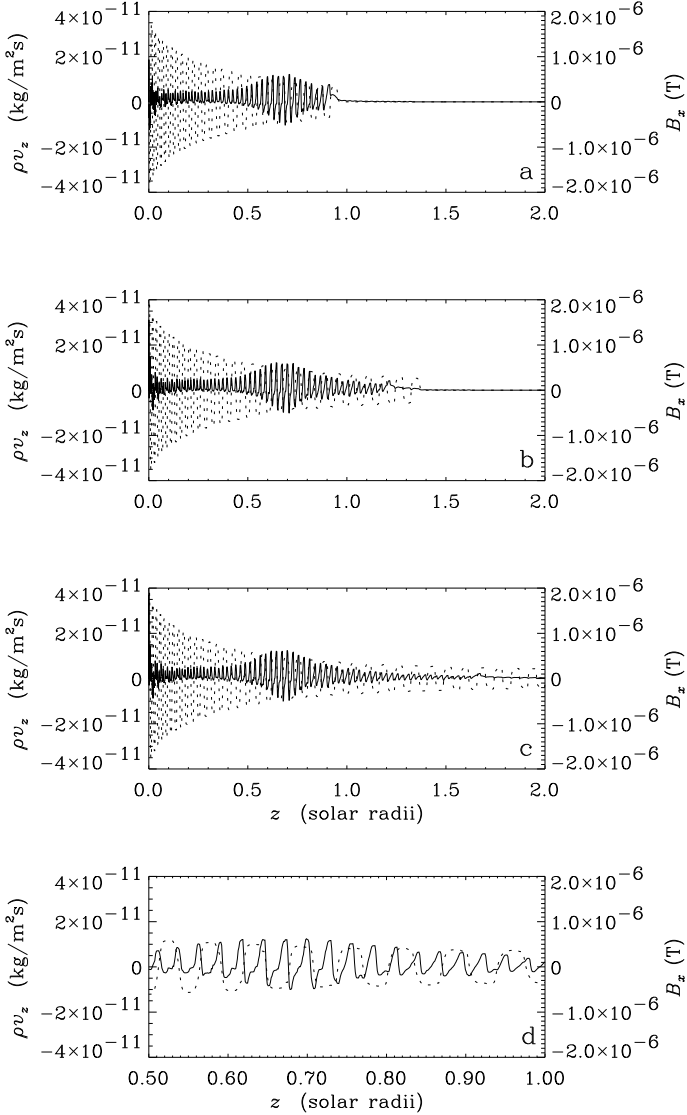}
\caption{ $\rho v_z$ (solid line/left scale) and 
$B_x$ (dotted line/right scale) for the Alfv\'en wave of Model 2b.  
({\bf a}), ({\bf b})
and ({\bf c}) are for the times 7\,200, 8\,800 and 11\,200 s, respectively.
({\bf d}) Shows an enlargement of the region between $0.5 R_{\sun}$ and 
$1 R_{\sun}$ in
({\bf c}).  Note how the acoustic oscillation (solid line) increases in 
amplitude up until $z = 0.7 R_{\sun}$, after which it decreases again, possibly
due to dissipation in shocks ({\bf d})}
\label{precursor_2}
\end{figure}

\begin{figure}
\epsfxsize=7.5cm
\epsfbox{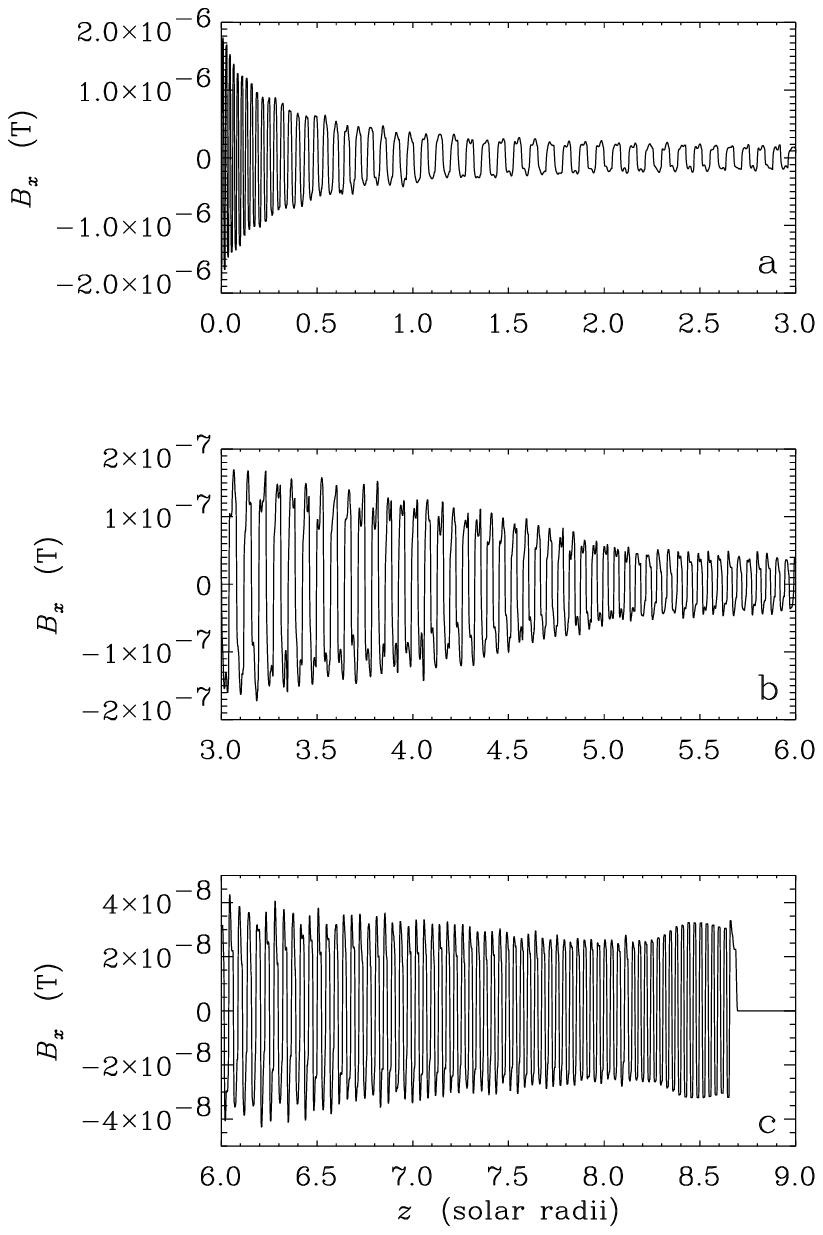}
\caption{$B_x$ as a function of $z$ for Model 2b at 50\,000 s.  ({\bf a})
$0 \le z \le 3 R_{\sun}$, ({\bf b}) $3 \le z \le 6 R_{\sun}$ and 
({\bf c}) $6 \le z \le 9 R_{\sun}$.  The Alfv\'en wave steepens to a square wave
({\bf a}), and loses energy in the current sheets ({\bf b}), but the first few
wavelengths in the head are essentially unaffected ({\bf c}).
Below $z = 1 R_{\sun}$, in particular,
there are signs of a low-frequency modulation of the
Alfv\'en wave}
\label{precursor_3}
\end{figure}

\subsubsection{Stratified models}

The stratified models are different with respect to the Alfv\'en velocity in
the sense that $v_{\rm A}$ first increases with $z$, but 
starts to
decrease again above $z = 2 R_{\sun}$.  Models 1a-c have $v_{\rm A} <
c_{\rm s}$ everywhere, so that we expect to see an acoustic precursor.
We increased the period to 900 s for these models, as a shorter wavelength
would have required a finer resolution, and thus more CPU time.
We show snapshots of Models 1b and c in Fig. \ref{precursor_6a}.  Figure 
\ref{precursor_6a}d shows clear signs of the nonlinear damping beyond $z = 1
R_{\sun}$ as $v_x$ stops increasing in amplitude.  The front
of the Alfv\'en wave has advanced further in Fig. \ref{precursor_6a}d than 
\ref{precursor_6a}b, which is surprising as both snapshots are taken after the
same simulation time.  It is the wave in Fig. \ref{precursor_6a}b that is
propagating at the expected Alfv\'en speed, whereas the nonlinear Alfv\'en wave
is propagating too fast (see Sect. \ref{super_fast_dis}).

Models 2a-2c are more interesting because the Alfv\'en speed
is comparable to the sound speed in the interval $1 R_{\sun} \le z \le 3
R_{\sun}$.  The most striking feature in the early stages of Model 2b is
however unrelated to this.  Figure \ref{precursor_2} shows how the acoustic
oscillations described by $\rho v_z$ grow in amplitude, but reach a 
maximum at $z = 0.7$.  After this the sound waves steepen and dissipate the
momentum that
they are carrying (Fig. \ref{precursor_2}d).  At 50\,000 s we see that
the amplitude of the Alfv\'en wave decreases sharply between $z = 3 R_{\sun}$
and $z = 5 R_{\sun}$
(Fig. \ref{precursor_3}b).
There are also signs of oscillations at higher frequencies superposed on
the wave, in particular downward of $z = 5 R_{\sun}$, and a general 
low-frequency modulation at $z < 1 R_{\sun}$.
Fig. \ref{precursor_3}c suggests that the first 7 or so wavelengths
of the wave are less damped and show less extraneous oscillations than the 
following wavelengths.
The same effects are present in Model 2c too, but in addition the wavefront
of Model 2c is propagating super-Alfv\'enically, which we will try to
explain in Sect. \ref{super_fast_dis}.  Model 2c crashed eventually
as the magnetic pressure evacuated a part
of the grid, which was expected as the pressure of the
oscillatory magnetic field is comparable to the gas pressure.

\begin{figure}
\epsfxsize=8.8cm
\epsfbox{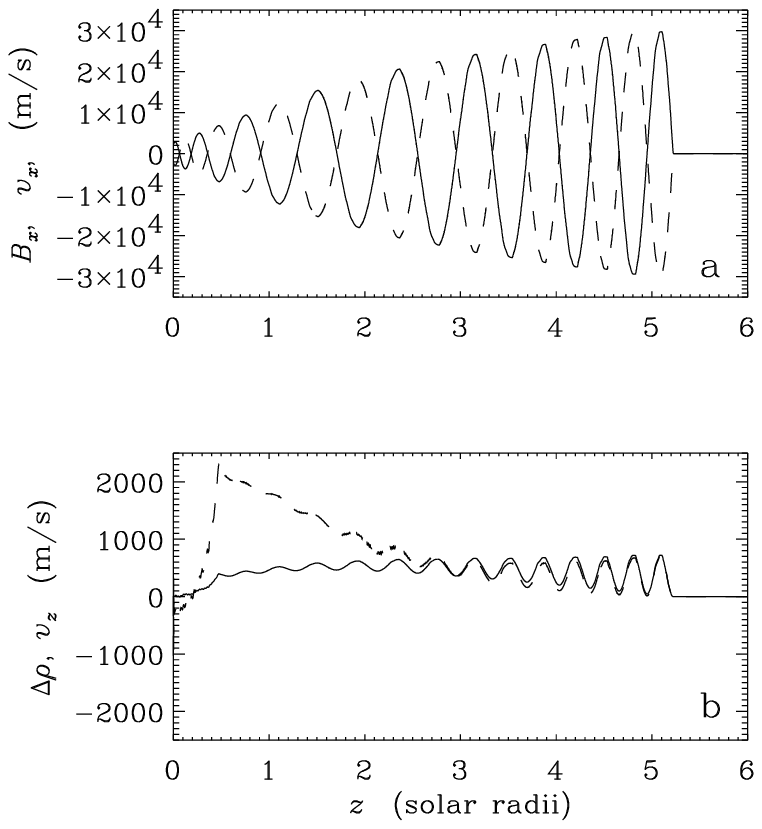}
\caption{A snapshot of ({\bf a}) $B_x$ (solid line) and $v_x$ (dashed line), 
and ({\bf b}) $v_z$ (solid line) and $\Delta \rho$
(dashed line)
at 2\,500 s for Model 3a.  
Note that that $B_x$ and $\Delta \rho$ are measured in velocity 
units by multiplying with $1/\sqrt{\mu_0\rho_0}$ and $c_{\rm s}/\rho_0$,
respectively.  
({\bf a}) shows the typical behaviour of an Alfv\'en wave, while
({\bf b}) shows density fluctuations carried by the Alfv\'en wave at
$z > 2.5 R_{\sun}$ 
and an outflow at smaller $z$}
\label{precursor_5}
\end{figure}

\begin{figure}
\epsfxsize=8.8cm
\epsfbox{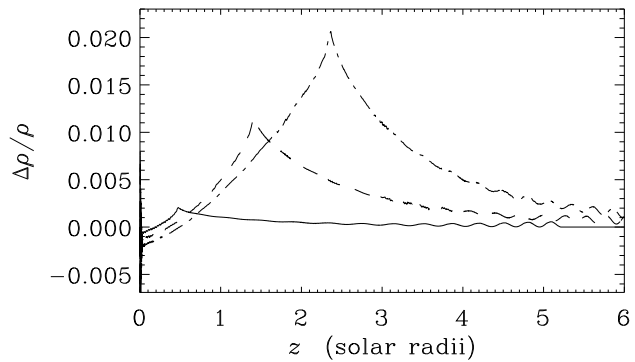}
\caption{$\Delta \rho/\rho_0$ at 2\,500 (solid line), 7\,500 (dashed 
line) and 12\,500 s (dot-dashed line) in Model 3a.  The peak of 
$\Delta \rho/\rho$ is moving at the isothermal sound speed}
\label{density}
\end{figure}

The only one of Models 3 that has got $(B_x^2+B_y^2)/(2\mu_0)$ significantly 
weaker
than the gas pressure is Model 3a.
There are no acoustic
precursors to the Alfv\'en waves in these models as $v_{\rm A} > c_{\rm s}$
everywhere, and indeed the most advanced parts of the density 
oscillations are not sound waves, but density 
fluctuations carried by the
Alfv\'en wave at the Alfv\'en speed.
Such a density fluctuation is expected to obey the relationship
\begin{equation}
  \frac{\Delta \rho}{\rho} = \frac{v_z}{v_{\rm A}}
\end{equation}
(Hollweg 1971, Eq. (15)).  At 2\,500 s this relationship is satisfied 
for $z > 2.5 R_{\sun}$, but at smaller heights there is a large density excess
(Fig. \ref{precursor_5}).  The density enhancement grows in extent 
with time (Fig. \ref{density}).  The local maximum is
moving outwards at the sound speed, $1.3\,10^5$ m\,s$^{-1}$.
All the Models 3 crash eventually, Model 3a and b at the time that they hit the
upper boundary, and Model 3c at a much earlier time, when the Alfv\'en wave
evacuates a part of the grid.

\begin{figure}
\epsfxsize=8.8cm
\epsfbox{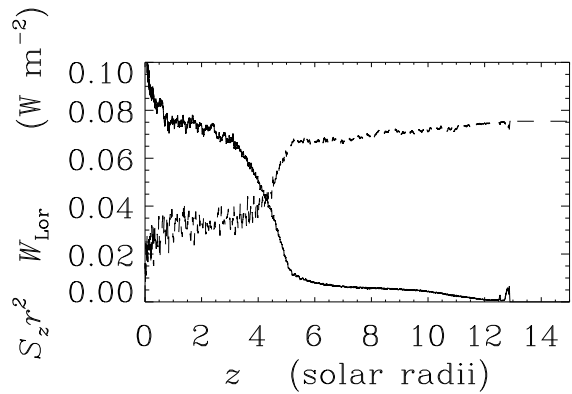}
\caption{$S_z \left(\frac{R+z}{R}\right)^2$ (solid line) and the integral
of the work done by the Lorentz force (dashed line) averaged over 300 s 
(the period of the Alfv\'en wave) 
for Model 2b}
\label{poynt_sphere}
\end{figure}

\begin{figure}
\epsfxsize=8.8cm
\epsfbox{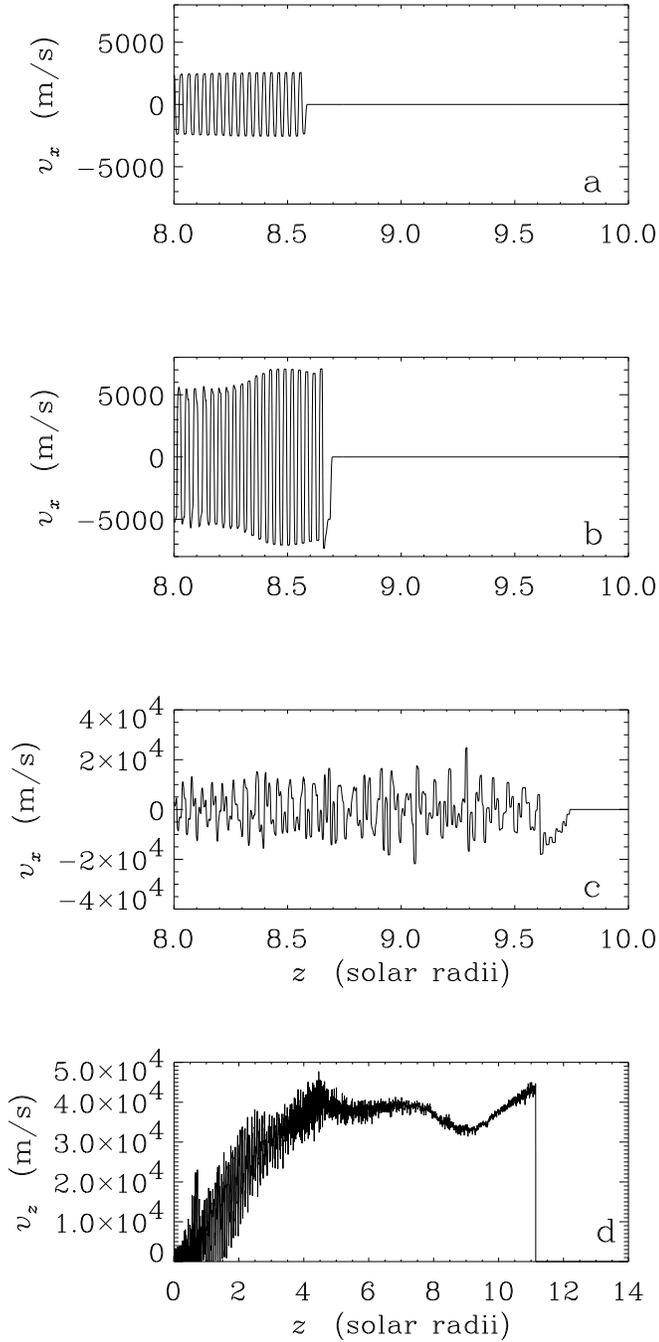}
\caption{$v_x$ for Models 2a-c at 50\,000 s ({\bf a} - {\bf c}), and $v_z$ for 
Model 2c ({\bf d}).  Note the change of $v_x$-scale between ({\bf b}) and 
({\bf c}), and the change of $z$-scale between ({\bf c}) and ({\bf d}).
This Fig. shows
that the wave front propagates faster the higher the amplitude of the wave}
\label{super_alfven}
\end{figure}

\subsection{Energetics}

As in Paper 1 the equations for the magnetic and kinetic energies are
\begin{equation}
  \frac{\partial}{\partial t} \frac{{\bmath B}^2}{2 \mu_0} + \nabla \cdot
{\bmath S}
= - \frac{{\bmath J}^2}{\sigma} - {\bmath v} \cdot \left({\bmath J \times 
\bmath B}\right),
\label{mag_enr}
\end{equation}
and
\begin{equation}
  \frac{\partial}{\partial t} \left( \frac{1}{2} \rho {\bmath v}^2\right) +
\nabla \cdot \left(\frac{1}{2} \rho {\bmath v}^2 {\bmath v} \right) = 
- {\bmath v} \cdot \nabla {\bmath p} + {\bmath v} \cdot \left({\bmath J 
\times \bmath B}\right)
+ \rho {\bmath v \cdot \bmath g},
\label{mech_enr}
\end{equation}
where ${\bmath S} = {\bmath E \times \bmath B}/\mu_0$ is the Poynting vector, 
and the
electric field is given by ${\bmath E} = - {\bmath v \times \bmath B} + 
{\bmath J}/\sigma$.
The Ohmic dissipation ${\bf J}^2/\sigma$ in Eq. (\ref{mag_enr}) represents 
the effect of the numerical diffusion in our code.  The numerical diffusion 
smears out discontinuities over the length scale of the grid spacing, but it is
at the same time negligible in regions with smooth velocities and magnetic
fields.  The thickness of a current sheet is proportional to $1/\sigma$, from
which it follows that ${\bf J} \propto \sigma$, so that the dissipation 
integrated over the current sheet is independent of $\sigma$.  The magnitude
of the numerical
diffusion should thus not affect the net damping unless the diffusion is strong
enough to damp out smooth variations, which is not the case.
As was pointed out in Paper 1 it is impossible to calculate the Ohmic 
dissipation directly from  the quantities of the numerical simulation,
however the other terms in Eq. (\ref{mag_enr}) are accessible.
We average these terms over the period of
the Alfv\'en wave.  
The averages of the nonlinear models are sensitive to secular changes and
aperiodic fluctuations in the simulations, which introduce some uncertainties
in the calculated energy losses.  We plot the time average of Model 2b as
an illustrative example in Fig. \ref{poynt_sphere}.  
10 - 20 \% of the Poynting flux is dissipated 
immediately
at the lower boundary due to imperfections in the boundary conditions and 
initial state,
and it may therefore be advisable to think of the waves as being of
correspondingly lower amplitude than indicated from Tab. 1.
Most of the remaining Poynting flux is lost between 3 and 5 $R_{\sun}$
by doing work on
the background medium via the Lorentz force.
The damping becomes more efficient with increasing amplitude of the Alfv\'en
wave both in the sense that more of the flux is lost from the wave and in the
sense that the damping sets in earlier (Tab. \ref{en_tab}).  
In addition a larger part of the
Poynting flux is spent on doing mechanical work instead of being lost through
Joule dissipation as the amplitude increases.

\begin{table*}
\caption{A compilation of the energetics of the Alfv\'en waves.  For every Model
the Poynting flux is given as the equivalent Poynting flux at the stellar
surface, $S_z \left(\frac{R+z}{R}\right)^2$, $S_z(0)$ is the Poynting flux
at $z = 0$, and $S_z({\rm final})$ the Poynting flux after the damping of 
the Alfv\'en wave.  We also give the work done by
the wave via the Lorentz force $W_{\rm Lor}$ per unit surface area and time, 
and the position where
half of the damping has taken place, $z_{\rm damp}$}
\begin{tabular}{lllll}
\hline
Model & $S_z(0)$ (W\,m$^{-2}$) & $S_z({\rm final})$
(W\,m$^{-2}$) & $W_{\rm Lor}$ (W\,m$^{-2}$) & $z_{\rm damp}\,\,(R_{\sun})$ \\
\hline
1a & $4\,10^{-5}$ & $3\,10^{-5}$ & $0.4\,10^{-5}$ & 8.2 \\
1b & $4\,10^{-3}$ & $3\,10^{-3}$ & $1\,10^{-3}$ & 7.2 \\
1c & $4\,10^{-1}$ & $1\,10^{-1}$ & $3\,10^{-1}$ & 1.4 \\
2a & $9\,10^{-4}$ & $0.4\,10^{-4}$ & $4\,10^{-4}$ & 9.3 \\
2b & $9\,10^{-2}$ & $0.5\,10^{-2}$ & $4\,10^{-2}$ & 4.4 \\
2c & 9 & .2 & 9 & 0.3 \\
\hline
\end{tabular}
\label{en_tab}
\end{table*}


\section{Discussion}

The most important and interesting results from our simulations are
\begin{itemize}

\item  The fact that practically all our Alfv\'en waves lose a significant
fraction of their Poynting flux within less than 10 $R_{\sun}$

\item  The fact that a nonlinear Alfv\'en wave can propagate 
super-Alfv\'enically

\item  The appearance of oscillations at lower frequencies than that at which
the Alfv\'en wave is driven

\end{itemize}
We will discuss each of these facts below, and also the significance of our
results for stellar wind models.

\subsection{The physical mechanism of the wave damping}

It was suggested in Paper 1 that the damping of the Alfv\'en waves takes place 
in current sheets.  These current sheets appear at the nodes of the
Alfv\'en wave, because the nodes represent minima of magnetic pressure, and
thus oppositely directed magnetic field lines are pushed together at the 
nodes.  This mechanism is
still at work in our new simulations, but at the same time, due to the 
stratification of the medium, there is a gradient in the wave pressure, which
in 
itself may do work on the medium (cf. Jacques 1977).  
In the nonlinear regime these effects are coupled together and cannot be
separated easily.
Table \ref{en_tab} shows that the Alfv\'en wave damping is 
a nonlinear effect as the fraction of the Poynting flux that is lost increases
with increasing initial Poynting flux.  In particular it is interesting to
find that $W_{\rm Lor}$ increases faster than $S_z(0)$, which we interpret as
the transverse magnetic field becoming more dynamically important compared to
the gas pressure.  The fact that $z_{\rm damp}$ decreases with $S_z(0)$ shows
that the damping cannot be described by a linear model.

\subsection{Super-Alfv\'enic motion}

\label{super_fast_dis}

We have seen in some of our simulations that nonlinear Alfv\'en waves can in
some circumstances propagate faster than the local Alfv\'en speed
(Fig. \ref{super_alfven}a-c).
The reason that the wave in Model 2c
can propagate at a higher speed than those in Models 2a and b
is that the wave itself is accelerating an
outflow with a maximum velocity of $4.8\,10^4$ m\,s$^{-1}$ (Fig. 
\ref{super_alfven}d).  Thus the Alfv\'en wave gains the velocity of the medium
it is propagating in.  This effect was not found in Paper 1, as in the 
plane-parallel models the Alfv\'en velocity is increasing upwards to eventually
become supersonic.  Consequently 
the Alfv\'en wave front manages to stay in front of
the main outflow (cf. Paper 1, Fig. 7), which is generated by the nonlinear 
part of the wave at lower altitudes.

\subsection{Low-frequency modulations}

\begin{figure*}
\epsfxsize=18cm
\epsfbox{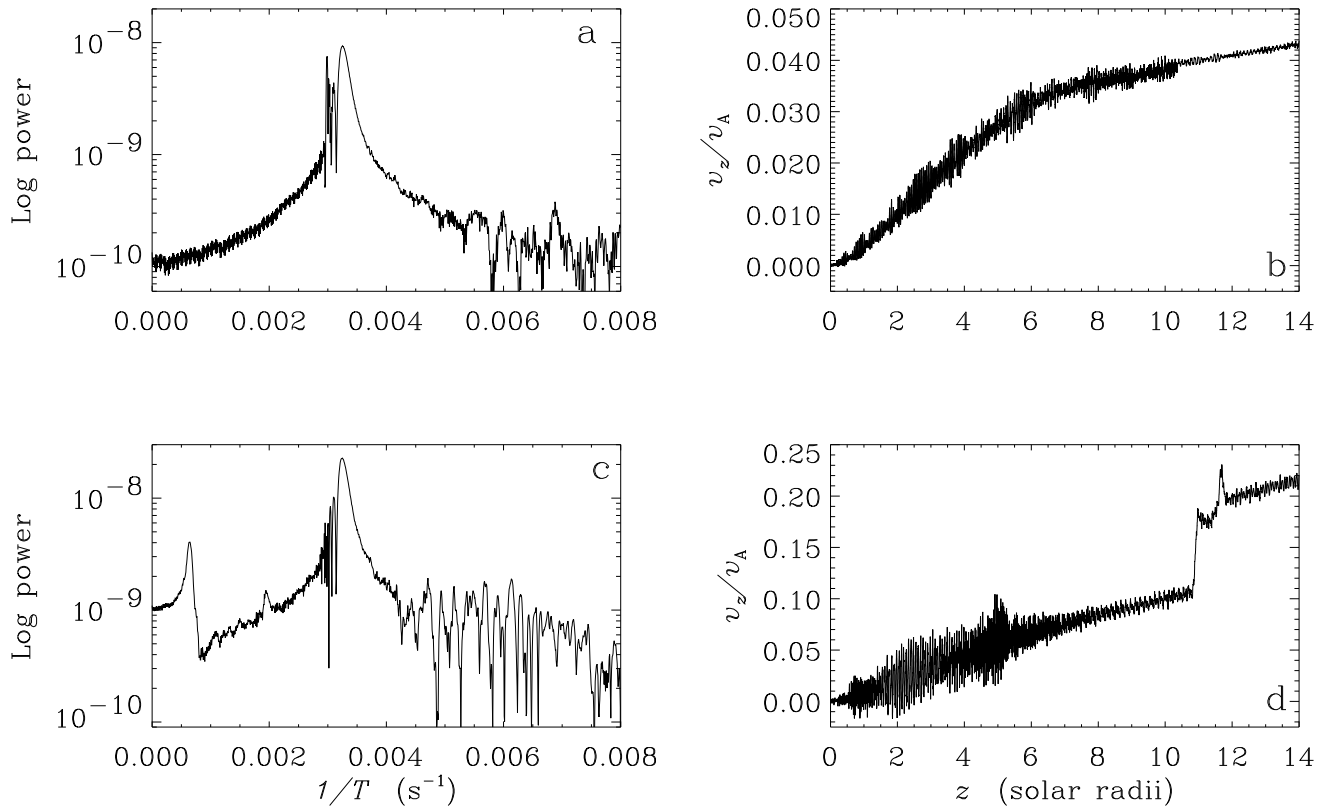}
\caption{Power spectra of $B_x$ for Models 1b ({\bf a}) and 2b ({\bf c}) 
at $t = 100\,000$ s.  ({\bf b}) and ({\bf d}) show $v_z$ in units of the
local Alfv\'en velocity at the same time for Models 1b and 2b, respectively}
\label{power}
\end{figure*}

As already noticed there is some evidence in Fig. \ref{precursor_3} that the
Alfv\'en wave is modulated on a frequency lower than the driving frequency.  To
test this we calculate power spectra of Models 1b and 2b 
(Fig. \ref{power}).  In addition to the main peak at 0.003\,2 Hz we see side 
peaks at 0.003\,0 and 0.000\,6 Hz.
To explain these extra peaks we note that a wave reflected back down by an
inhomogeneity propagating upwards is Doppler-shifted to a lower frequency
\begin{equation}
  \nu = \nu\left(0\right) \frac{v_{\rm A} - v_z}{v_{\rm A} + v_z},
\end{equation}
where $\nu\left(0\right)$ is the original frequency of the wave.  To explain the side peak
in Model 1b we require an outflow velocity $v_z = 0.04 v_{\rm A}$, which
agrees well with the velocities generated in the model (Fig. \ref{power}b).
The peak at 0.000\,6 Hz of Model 2b on the other hand requires $v_z = 0.7 
v_{\rm A}$, which is significantly higher than what is available (Fig.
\ref{power}d).  An alternative interpretation in this case is that the
frequency 0.000\,6 Hz represents the beat frequency between the original 
and reflected Alfv\'en waves, thus giving the frequency of the reflected 
Alfv\'en wave as 0.002\,6 Hz, corresponding to $v_z = 0.1 v_{\rm A}$.  This
agrees well with the velocity before the jump in $v_z$ at 11 $R_{\sun}$ (Fig.
\ref{power}d), but the peak
in the power spectrum (Fig. \ref{power}c) is at 0.002 Hz, which may be a 
consequence of the inhomogeneity of the medium.

\subsection{Importance for stellar winds}

Alfv\'en-wave driven winds have been discussed several times in the past
(e.g. Hartmann \& MacGregor 1980, 1982; Leer et al. 1982, 
MacGregor \& Charbonneau 1994).
The main reason is that the Alfv\'en waves may be able to drive the wind
in circumstances where more well-understood mechanisms can be shown to be
insufficient.  This is for instance the case for the fast solar wind moving
at velocities of 800 km\,s$^{-1}$, which is too fast for the classical thermally
driven wind model by Parker (1958), and the winds of late-type giants, where
the temperatures are too low to drive a wind.  A severe weakness of the
models of the winds of late-type stars is that the Alfv\'en waves must be
damped within a few stellar radii in order to avoid over-accelerating
the wind.  Usually this has been done by ascribing an arbitrary damping length
to the Alfv\'en waves, but the resulting models are sensitive to the choice
of the damping length (Holzer et al. 1983).  
In our simulations the waves
are damped without the action of any dissipative mechanism while they are 
doing work on the background medium.  
Our mechanism does not 
require the wavelength to become comparable to the length scale of the Alfv\'en
velocity variations in contrast to some previous attempts that have appealed
to non-WKB waves being reflected.  Although we have not surveyed the parameter
space it appears likely that the parameters of
our model can be modified to provide a
sufficiently effective damping to fulfill the constraints put on those
wind models.  An obvious improvement on our current simulations is to start from
an initial state which describes an outflowing wind.

Magnetohydrodynamic waves may have been observed in the inner parts of the 
solar wind.  Ofman \& Davila (1997b) have calculated that magnetohydrodynamic
waves
in the solar wind may give a line broadening of $\sim 300$ km\,s$^{-1}$,
which is comparable to the line widths observed by {\it SOHO} (e.g. Kohl et al.
1996).  In particular
Ofman \& Davila (1997a) suggests that what they call {\em solitary}
waves are essential for the acceleration of the solar wind.  The solitary waves
are propagating $\rho$-$v_z$-perturbations generated by
Alfv\'en waves, and thus are akin to the density oscillations we have found
in for instance Fig. \ref{precursor_2}.  The authors call them solitary
waves as they believe them to be related to the solitary wave solutions
found in slabs and thin flux tubes by Roberts (1981) and Roberts \& Mangeney
(1982).  This is an interpretation that we find questionable for the following
reason.  It is certainly true that the two-dimensional axisymmetric model 
used by Ofman \& Davila bears some resemblance to a flux tube with a lower
density in the interior, but our one-dimensional simulations lack all features
characteristic of a flux tube, and still we find the same kind of density
oscillations propagating faster than the sound speed.  We thus conclude that
the oscillations cannot be related to the existence of solitons in flux tubes.
Regarding the question of whether the solar wind is driven by Alfv\'en 
waves or {\em solitary} waves that is very much a matter of semantics as the
{\em solitary} waves must be driven by the Alfv\'en waves.  It is interesting
to compare the simulations in terms of efficiency of deriving energy from the
Alfv\'en waves.  The background models are roughly similar apart from that
Ofman \& Davila assume a radial magnetic field which is 70\% stronger than the
one we use in Models 3.  They drive the Alfv\'en wave at an amplitude
inbetween our Models 3a and b, and assume a period which
is almost an order of magnitude longer than ours.  Consequently they find 
larger radial velocities than we do, but unfortunately they do not give any
numbers for the fraction of the Poynting flux that has been converted to 
mechanical energy.  They do however state that $3\,10^{-3}$ of the Poynting 
flux has been lost due to Ohmic dissipation.  This appears to be a rather 
inefficient conversion process, but one must keep in mind that
their
grid extends only to 4 $R_{\sun}$, and we typically find that most of the 
damping takes place at larger distances. 

Naturally there are alternative models for driving the fast solar wind.  
Feldman et al. (1996) have suggested that the outflows are driven by the
same kind of reconnection events that produce the X-ray jets (Yokoyama \&
Shibata 1995).  In this model the emerging magnetic field of a bipolar region
collides with the magnetic field of the chromospheric
network.  This field is
concentrated in narrow flux tubes in the photosphere, but due to the 
stratification the flux tube increases in radius upwards.  At the collision
point the magnetic fields reconnect and
two jets moving at close to the 
Alfv\'en velocity appear.  Only a part of the momentum of the jet is 
directed upwards,
but the horizontal momentum is absorbed by nearby matter and magnetic field,
so that also a part of this becomes available for generating an outflow from
the coronal hole.  Inevitably some of the energy goes into producing Alfv\'en
waves, and so the model may also explain the presence of Alfv\'en waves in
the solar wind.

\section{Conclusions}

In this paper we have studied the propagation of nonlinear spherical Alfv\'en
waves.  Like in our previous simulations of Alfv\'en waves in a
plane-parallel atmosphere the waves damp by forming current sheets in which 
Poynting
flux is lost to Ohmic heating and the acceleration of an outflow.
In general most of the Poynting flux is spent
on accelerating an outflow.  
This combined process
of damping Alfv\'en waves and accelerating an outflow may be important in 
understanding both the fast solar wind and
the winds of late-type giants.

\section*{Acknowledgements}
Our work on nonlinear Alfv\'en waves was begun as
part of a collaborative research project between the Astronomical Institute,
Utrecht, and the FOM Institute for Plasma Physics, Rijnhuizen, the Netherlands.
We are grateful to the other members of this collaboration, J. P. Goedbloed,
A. G. Hearn, S. Poedts and G. T\'oth for a stimulating environment.  UT is
supported by an EU post-doctoral fellowship.  We thank G. Ogilvie for reading 
the manuscript and providing useful comments, and J. Orta for spotting an
error in our code.  Finally we thank an anonymous referee for helpful comments.


\begin{thebibliography}{}

\bibitem[\protect\citename{An et al. }1990]{an:etal}
An, C.-H., et al., 1990,
ApJ, 350, 309

\bibitem[\protect\citename{Balogh et al. }1995]{balogh}
Balogh A., et al., 1995,
Sci, 268, 1007

\bibitem[\protect\citename{Belcher \& Davis }1971]{belcher:davis}
Belcher, J. W., Davis Jr., L., 1971, J. Geophys. Res., 76, 3534

\bibitem[\protect\citename{Bohlin }1976]{bohlin}
Bohlin, J. D., 1976,
in D. J. Williams (ed.)  Physics of solar planetary environments, American 
Geophysical Union, 114

\bibitem[\protect\citename{Boynton \& Torkelsson }1996]{boynton:torkelsson}
Boynton G. C., Torkelsson U., 1996, A\&A, 308, 299 (Paper 1)

\bibitem[\protect\citename{Feldman }1996]{feldman}
Feldman, W. C., et al., 1996,
A\&A, 316, 355

\bibitem[\protect\citename{Hartmann \& MacGregor }1980]{hartmann:macgregora}
Hartmann, L., MacGregor, K. B., 1980, ApJ, 242, 260

\bibitem[\protect\citename{Hartmann \& MacGregor }1982]{hartmann:macgregorb}
Hartmann, L., MacGregor, K. B., 1982, ApJ, 257, 264

\bibitem[\protect\citename{Heinemann \& Olbert }1980]{heinemann:olbert}
Heinemann, M., Olbert, S., 1980, J. Geophys. Res., 85, 1311

\bibitem[\protect\citename{Hollweg }1971]{hollweg}
Hollweg, J. V., 1971, J. Geophys. Res., 76, 5155

\bibitem[\protect\citename{Holzer et al. }1983]{holzer}
Holzer, T. E., Fl\aa , T., Leer, E., 1983, ApJ, 275, 808

\bibitem[\protect\citename{Jacques }1977]{jacques}
Jacques, S. A., 1977, ApJ, 215, 942

\bibitem[\protect\citename{Kohl }1996]{kohl}
Kohl, J. L., et al., 1996,
BAAS, 28, 897

\bibitem[\protect\citename{Lamb }1908]{lamb}
Lamb, H., 1908, Proc. Lond. Math. Soc., (2), VII, 122

\bibitem[\protect\citename{Lamb }1932]{lamb2}
Lamb, H., 1932, Hydrodynamics, Cambridge University Press, Cambridge

\bibitem[\protect\citename{Landau \& Lifshitz }1987]{landau:lifshitz}
Landau, L. D., Lifshitz, E. M., 1987,
Fluid Mechanics,
Pergamon Press, Oxford

\bibitem[\protect\citename{Leer et al. }1982]{leer:etal}
Leer, E., Holzer, T. E., Fl\aa , T., 1982, Space Sci. Rev., 33, 161

\bibitem[\protect\citename{Lou \& Rosner }1994]{lou:rosner}
Lou, Y.-Q., Rosner, R., 1994,
ApJ, 424, 429

\bibitem[\protect\citename{MacGregor \& Charbonneau }1994]
{macgregor:charbonneau}
MacGregor, K. B., Charbonneau, P., 1994, ApJ, 430, 387

\bibitem[\protect\citename{Ofman \& Davila }1997a]{ofman:davilaa}
Ofman, L., Davila, J. M., 1997a,
ApJ, 476, 357

\bibitem[\protect\citename{Ofman \& Davila }1997b]{ofman:davilab}
Ofman, L., Davila, J. M., 1997b,
ApJ, 476, L51

\bibitem[\protect\citename{Parker }1958]{parker}
Parker, E. N., 1958, ApJ, 128, 664

\bibitem[\protect\citename{Roberts }1981]{roberts}
Roberts, B., 1981,
Solar Phys., 69, 39

\bibitem[\protect\citename{Roberts \& Mangeney }1982]{roberts:mangeney}
Roberts, B., Mangeney, A., 1982,
MNRAS, 198, 7P

\bibitem[\protect\citename{Yokoyama \& Shibata }1995]{yokoyama:shibata}
Yokoyama, T., Shibata, K., 1995,
Nat, 375, 42

\bibitem[\protect\citename{Zirker }1977]{zirker}
Zirker, J. B., 1977,
Coronal holes and high speed wind streams, Colorado University Press, Boulder

\end{thebibliography}
\end{document}